\documentclass[12pt,english,journal,12pt,onecolumn]{IEEEtran}
\usepackage[T1]{fontenc}
\usepackage[latin9]{inputenc}
\usepackage{geometry}
\usepackage{units}
\usepackage{amsmath}
\usepackage{amssymb}
\usepackage{graphicx}

\makeatletter

\usepackage{amsfonts}

\makeatother

\usepackage{babel}
\begin{document}

\title{Medium Access Control Protocols for Wireless Sensor Networks with
Energy Harvesting}

\author{F. Iannello, O. Simeone and U. Spagnolini%
\thanks{F. Iannello is with both Politecnico di Milano, Milan, 20133, Italy
and the Center for Wireless Communications and Signal Processing Research
(CWCSPR), New Jersey Institute of Technology (NJIT), Newark, New Jersey
07102-1982 USA (e-mail: iannello@elet.polimi.it). U. Spagnolini is
with Politecnico di Milano. O. Simeone is with the CWCSPR, NJIT.%
}}
\maketitle
\begin{abstract}
The design of Medium Access Control (MAC) protocols for wireless sensor
networks (WSNs) has been conventionally tackled by assuming battery-powered
devices and by adopting the network lifetime as the main performance
criterion. While WSNs operated by energy-harvesting (EH) devices are
not limited by network lifetime, they pose new design challenges due
to the uncertain amount of harvestable energy. Novel design criteria
are thus required to capture the trade-offs between the potentially
infinite network lifetime and the uncertain energy availability.

This paper addresses the analysis and design of WSNs with EH devices
by focusing on conventional MAC protocols, namely TDMA, Framed-ALOHA
(FA) and Dynamic-FA (DFA), and by accounting for the performance trade-offs
and design issues arising due to EH. A novel metric, referred to as
\textit{delivery probability}, is introduced to measure the capability
of a MAC protocol to deliver the measure of any sensor in the network
to the intended destination (or \textit{fusion center}, FC). The interplay
between delivery efficiency and \textit{time efficiency} (i.e., the
data collection rate at the FC), is investigated analytically using
Markov models. Numerical results validate the analysis and emphasize
the critical importance of accounting for both delivery probability
and time efficiency in the design of EH-WSNs.\end{abstract}
\begin{IEEEkeywords}
Wireless sensor networks, multiaccess communication, energy harvesting,
dynamic framed ALOHA.
\end{IEEEkeywords}

\section{Introduction}

Recent advances in low-power electronics and energy-harvesting (EH)
technologies enable the design of self-sustained devices that collect
part, or all, of the needed energy from the surrounding environment.
Several systems can take advantage of EH technologies, ranging from
portable devices to wireless sensor networks (WSNs) \cite{art: ParadisoStarner}.
However, EH devices open new design issues that are different from
conventional battery-powered (BP) systems \cite{art: Kansal}, where
the main concern is the network lifetime \cite{art: Akyildiz}. In
fact, EH potentially allows for perpetual operation of the network,
but it might not guarantee short-term activities due to temporary
energy shortages \cite{art: Kansal}. This calls for the development
of energy management techniques tailored to the EH dynamics. While
such techniques have been mostly studied at a single-device level
\cite{art: Sharma}, in wireless scenarios where multiple EH devices
interact with each other, the design of EH-aware solutions needs to
account for a system-level approach \cite{art: Koksal}\cite{art: SharmaMAC}.
This is the motivation of this work.

\begin{figure}[h!]
\centering \includegraphics[clip,width=10cm]{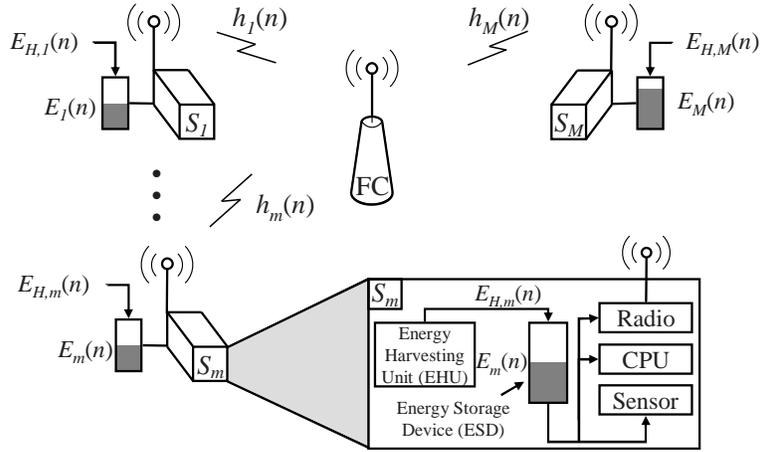} \caption{WSN with a single Fusion Center (FC) gathering data from $M$ sensors,
which are equipped with an energy storage device (ESD) and an energy-harvesting
unit (EHU).}

\label{fig:system_model}
\end{figure}

In this paper, we focus on system-level design considerations for
WSNs operated by EH-capable devices. In particular, we address the
analysis and design of medium access control (MAC) protocols for single-hop
WSNs (see Fig. \ref{fig:system_model}) where a \textit{fusion center}
(FC) collects data from sensors in its surrounding. Specifically,
we investigate how performance and design of MAC protocols routinely
used in WSNs, such as TDMA \cite{Lib: Bertsekas}, Framed-ALOHA (FA)
and Dynamic-FA (DFA) \cite{art: Schoute}, are influenced by the discontinuous
energy availability in EH-powered devices.

\subsection{State of the Art}

In recent years, WSNs with EH-capable nodes have been attracting a
lot of attention, also at commercial level. To provide some examples,
the Enocean Alliance proposes to use a MAC protocol for EH devices
based on pure ALOHA strategies \cite{Ind.: EnOcean}, while an enhanced
self-powered RFID tag created by Intel, referred to as WISP \cite{art: WISP},
has been conceived to work with the EPC Gen 2 standard \cite{std: EPC-standard}
that adopts a FA-like MAC protocol.

However, while performance analysis of MAC protocols in BP-WSNs have
been investigated in depth (see e.g., \cite{Lib: Bertsekas}\cite{art: Schoute}\cite{art: Ephremedis}),
analyses of MAC protocols with EH are hardly available. A notable
exception is \cite{art: SharmaMAC}, where data queue stability has
been studied for TDMA and carrier sense multiple access (CSMA) protocols
in EH networks. We remark that routing for EH networks has instead
received more attention, see e.g., \cite{art: Kansal}\cite{art:Tassiulas}.

\subsection{Contributions}

In this paper we consider the design and analysis of TDMA, FA and
DFA MAC protocols in the light of the novel challenges introduced
by EH. In Sec. \ref{Sec: MAC_protocols} we propose to measure the
system performance in terms of the trade-off between the \textit{delivery
probability}, which accounts for the number of sensors' measurements
successfully reported to the FC, and the \textit{time efficiency},
which measures the rate of data collection at the FC (formal definitions
are in Sec. \ref{Sec: MAC_protocols}). We then introduce an analytical
framework in Sec. \ref{Sec: Analysis_Performance_Metrics} and Sec.
\ref{Sec: ESDenergyEvolution} to assess the performance of EH-WSNs
in terms of the mentioned trade-off for TDMA, FA and DFA MAC protocols.
In Sec. \ref{Sec: BacklogEstimation} we tackle the critical issue
in ALOHA-based MAC protocols of estimating the number of EH sensors
involved in transmission, referred to as \textit{backlog}, by proposing
a practical reduced-complexity algorithm. Finally, we present extensive
numerical simulations in Sec. \ref{Sec: Numerical_Results} to get
insights into the MAC protocol design trade-offs, and to validate
the analytical derivations.

\section{System Model\label{Sec: SystemModel}}

In this paper, we consider a single-hop WSN with a FC surrounded by
$M$ wireless sensors labeled as $S_{1},S_{2},...,S_{M}$ (see Fig.
\ref{fig:system_model}). Each sensor (or \textit{user}) is equipped
with an EH unit (EHU) and an energy storage device (ESD), where the
latter is used to store the energy harvested by the EHU. The FC retrieves
measurements from sensors via periodic \textit{inventory rounds }(IRs),
once every $T_{int}$ seconds $[s]$. Each IR is started by the FC
by transmitting an initial\textit{ query command} (Q), which provides
both synchronization and instructions to sensors on how to access
the channel. Time is slotted, with each \textit{slot} lasting $T_{s}$
$[\unit{s}]$. The effective duration of the $n$th IR, during which
communication between the FC and the sensors takes place, is denoted
by $T_{IR}(n)$. We assume that $T_{IR}(n)\ll T_{int}$ for all IR
$n$, and also that the query duration is negligible, so that the
ratio $T_{IR}(n)/T_{s}$ indicates the total number of slots allocated
by the FC during the $n$th IR.

In every IR, each sensor has a new measure to transmit with probability
$\alpha$, independently of other sensors and previous IRs. If a new
measure is available, the sensor will mandatory attempt to report
it successfully to the FC as long as enough energy is stored in its
ESD (see Sec. \ref{Sec: Energy_model} for details). Each measure
is the payload of a packet, whose transmission fits within the slot
duration $T_{s}$. Sensors' transmissions within each IR are organized
into \textit{frames}, each of which is composed of a number of slots
that is selected by the FC. Depending on the adopted MAC protocol,
any user that needs to (and can) transmit in a frame either chooses
or is assigned a single slot within the frame for transmission as
it will be detailed below. Moreover, after a user has successfully
transmitted its packet to the FC, it first receives an acknowledge
(ACK) of negligible duration by the FC and then it becomes inactive
for the remaining of the IR. We emphasize that the FC knows neither
the number of sensors with a new measure to transmit, nor the state
of sensors' ESDs.

\subsection{Interference Model\label{Sec: Interference_Model}}

We consider \textit{interference-limited }communication scenarios
where the \textit{downlink} packets transmitted by the FC are always
correctly received (error-free) by the sensors, while \textit{uplink}
packets transmitted by the sensors to the FC are subject to communication
errors due to possible interference arising from collisions with other
transmitting sensors. The uplink channel power gain for the $m$th
sensor during the $n$th IR is $h_{m}(n)$. Channel gain $h_{m}(n)$
is assumed to be constant over the entire IR but subject to random
independent and identically distributed (i.i.d.) fading across IRs
and sensors, with pdf $f_{h}(\cdot)$ and normalized such that $E\left[h_{m}(n)\right]=1$,
for all $n,m$. In the presence of simultaneous transmissions within
the same slot during the $k$th frame of the $n$th IR, a sensor,
say $S_{m}$, is correctly received by the FC if and only if its instantaneous
signal-to-interference ratio (SIR) $\gamma_{m,k}\left(n\right)$ is
larger than a given threshold $\gamma_{th}$, i.e., if 
\begin{equation}
\gamma_{m,k}\left(n\right)=\frac{h_{m}\left(n\right)}{\sum_{l\in\mathcal{I}_{m,k}(n)}h_{l}\left(n\right)}\geq\gamma_{th},\label{eqn: Istantaneous_SIR}
\end{equation}
where $\mathcal{I}_{m,k}(n)$ denotes the set of sensors that transmit
in the same slot selected by $S_{m}$ in frame $k$ and IR $n$. We
assume $\gamma_{th}>0dB$ so that, in case a slot is selected by more
than one sensor, at most one of the colliding sensor can be successfully
decoded in the slot.

According to the interference model (\ref{eqn: Istantaneous_SIR}),
any slot can be: \textit{empty} when it is not selected by any sensor;
\textit{collided} when it is chosen by more than one sensors but none
of them transmits successfully; \textit{successful} when one sensor
transmits successfully possibly in the presence of other (interfering)
users. Successful transmission in the presence of interfering users
within the same slot is often referred to as \textit{capture effect
}\cite{art: Ephremedis}.

\textbf{Remark 1}: Errors in the decoding of downlink query packets
can be accounted for through the parameter $\alpha$ as well. In fact,
let $\alpha_{Q}$ be the probability that a user correctly decodes
the downlink packet sent by the FC at the beginning of an IR. Moreover,
assume that downlink decoding errors are i.i.d. across sensors and
IRs, and let $\alpha_{N}$ be the probability that a user has a new
measure to transmit in any IR. Then, the probability that any user
$S_{m}$ has a new measure and correctly decodes the FC's query is
given by the product $\alpha=\alpha_{Q}\alpha_{N}$.

\subsection{ESD and Energy Consumption Models\label{Sec: Energy_model}}

We consider a discrete ESD with $N+1$ energy levels in the set $\mathcal{E}=\{0,\mathcal{\delta},2\mathcal{\delta},...,N\mathcal{\delta}\}$,
where $\mathcal{\delta}$ is referred to as \textit{energy unit}.
Let $E_{m}(n)\in\mathcal{E}$ be the energy stored in the ESD of the
$m$th user at the beginning of the $n$th IR. Energy $E_{m}(n)$
is a random variable that is the result of the EH process and the
energy consumption of the sensor across IRs; its probability mass
function (pmf) is $p_{E(n)}\left(\cdot\right)$ and the corresponding
complementary cumulative distribution function (ccdf) is $G_{E(n)}\left(x\right)=\Pr[E_{m}(n)\geq x]$.
Note that, the initial energy distribution $p_{E(1)}\left(\cdot\right)$
is given, while the evolution of the pmf $p_{E(n)}\left(\cdot\right)$
for $n>1$ depends on both the MAC protocol and EH process.

\begin{figure}[h!]
\centering \includegraphics[clip,width=10cm]{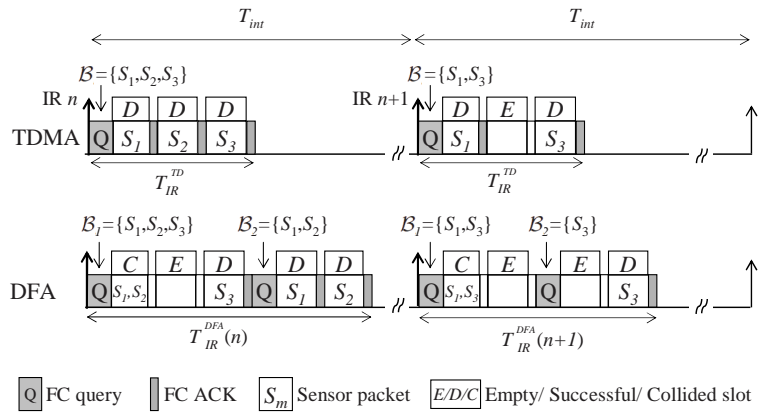}
\caption{Examples for TDMA and DFA MAC protocols for $M=3$. FA is not depicted
since it is a special case of DFA with only one frame. The backlog
for each frame is indicated above each query. Some sensors might not
be in the backlog due to energy shortage and/or absence of a new measure
to report.}

\label{fig:protocol_description}
\end{figure}

We assume that each time a sensor transmits a packet it consumes an
energy $\varepsilon$, which accounts for the energy consumed in the:
\textit{a}) reception of the FC's query that starts the frame (see
Fig. \ref{fig:protocol_description}); \textit{b}) transmission; \textit{c})
reception of FC's ACK or not ACK (NACK) packet. At the beginning of
each IR, a sensor with a new measure to transmit can participate to
the current IR only if the energy stored in its ESD is at least $\varepsilon$.
Let $\varepsilon_{\mathcal{\delta}}=\varepsilon/\mathcal{\delta}$
be the number of energy units $\mathcal{\delta}$ required for transmission,
where $\varepsilon_{\mathcal{\delta}}$ is assumed to be an integer
value without loss of generality. Let $F_{\varepsilon}=N\mathcal{\delta}/\varepsilon=N/\varepsilon_{\mathcal{\delta}}$
be the (normalized) \emph{capacity} of the ESD, which is assumed to
be an integer indicating the maximum number of (re)transmissions allowed
by a fully charged ESD.\newpage

\subsection{Energy Harvesting Model\label{Sec: Energy_Harvesting_model}}

During the time $T_{int}$ between the $n$th and $(n+1)$th IRs the
$m$th sensor $S_{m}$ harvests an energy $E_{H,m}(n)$, which is
modeled as a discrete random variable, i.i.d. over IRs and sensors,
with pmf $q_{i}=\Pr[E_{H,m}(n)=i\mathcal{\delta}]$, with $i\in\{0,1,2,...\}$,
and for all $m$ and $n$. For technical reasons that we discuss in
Sec. \ref{Sec: DMCmodel}, we assume that the probability $q_{0}$
and $q_{1}$ of harvesting zero and one energy unit respectively,
are both strictly positive, namely $q_{0}>0$ and $q_{1}>0$.

We assume that the EH dynamics is much slower than the IR duration
$T_{IR}(n)$, so that the amount of energy harvested within $T_{IR}(n)$
can be considered as negligible with respect to $\varepsilon$ (recall
also that $T_{IR}(n)\ll T_{int}$). Hence, the only energy that a
sensor can actually use throughout an IR is the energy initially available
at the beginning of the IR itself (i.e., $E_{m}(n)$).

\section{Performance Metrics and Medium Access Control Protocols\label{Sec: MAC_protocols}}

We first introduce in Sec. \ref{Sec: MACperformanceMetrics} the considered
performance metrics, namely delivery probability and time efficiency,
and then in Sec. \ref{Sec: MAC_protocols_description} we review the
considered MAC protocols.

\subsection{MAC Performance Metrics\label{Sec: MACperformanceMetrics}}

\subsubsection{Delivery Probability\label{Sec: AsymptoticDeliveryProbability}}

The delivery probability $p_{d}\left(n\right)$ measures the capability
of the MAC protocol to successfully deliver the measure of any sensor,
say $S_{m}$, to the FC during the $n$th IR 
\begin{equation}
p_{d}(n)=\Pr\left[S_{m}\text{ transmits successfully in IR }n|\text{ }S_{m}\text{ has a new measure in IR }n\right].\label{eqn: DeliveryProbability}
\end{equation}
The statistical equivalence of all sensors makes the probability (\ref{eqn: DeliveryProbability})
independent of the specific sensor. Notice that a sensor fails to
report its measure during an IR if either it has an energy shortage
before (re)transmitting the packet correctly, or the MAC protocol
does not provide the sensor with sufficient retransmission opportunities.
Given the potentially perpetual operation enabled by EH, it is relevant
to evaluate the delivery probability when the system is in \textit{steady-state}.
The \textit{asymptotic delivery probability} is thus obtained by taking
the limit of $p_{d}\left(n\right)$ for large IR index $n$, provided
that it exists, as 
\begin{equation}
p_{d}^{AS}=\underset{n\rightarrow\infty}{\lim}p_{d}(n).\label{eqn: Asymptotic_Delivery_Probability}
\end{equation}

\subsubsection{Time Efficiency\label{Sec: AsymptoticTimeEfficiency}}

The time efficiency $p_{t}(n)$ measures the probability that any
slot allocated\ by the MAC within the $n$th IR is successfully used
(i.e., it is neither empty nor collided, see Sec. \ref{Sec: Interference_Model})
\begin{equation}
p_{t}\left(n\right)=\Pr\left[\text{The FC correctly retrieves a packet in any slot of the }n\text{th IR}\right].\label{eqn: Time_efficiency}
\end{equation}
By taking the limit of (\ref{eqn: Time_efficiency}) for $n\rightarrow\infty$,
we obtain the\textit{ asymptotic time efficiency} 
\begin{equation}
p_{t}^{AS}=\underset{n\rightarrow\infty}{\lim}p_{t}\left(n\right).\label{eqn: Asymptotic_Time_efficiency}
\end{equation}

\textbf{Remark 2:} Informally speaking, the time efficiency $p_{t}(n)$
measures the ratio between the total number of packets successfully
received by the FC and the total number of slots allocated by the
MAC protocol (i.e., $T_{IR}(n)/T_{s}$, see Sec. \ref{Sec: SystemModel}).
As it will be shown in Sec. \ref{Sec: MAC_protocols_description},
the IR duration $T_{IR}(n)$ is in general a random variable, and
consequently, time efficiency $p_{t}(n)$ differs from more conventional
definitions of throughput (see e.g., \cite{art: Schoute}) which measure
the number of packets delivered over the interval between two successive
IRs $T_{int}$, instead of $T_{IR}(n)$. The rationale for this definition
of time efficiency is that it actually captures more effectively the
rate of data collection at the FC. Whereas, the delivery probability
accounts for the fraction of users, with a new measure to transmit
at the beginning of the current IR, which are able to successfully
report their payload to the FC within the IR, where delivery failures
are due to collisions and energy shortages.

In contention based MACs (e.g., ALOHA), there is a trade-off between
delivery probability and time efficiency. In fact, increasing the
former generally requires the FC to allocate a larger number of slots
in an IR to reduce packet collisions, which in turn decreases the
time efficiency.

\subsection{MAC Protocols\label{Sec: MAC_protocols_description}}

In this section, we review the standard MAC protocols that we focus
on.

\subsubsection{TDMA\label{Sec: TDMAprotocol}}

With the TDMA protocol, each user is pre-assigned an exclusive slot
that it can use in every IR, irrespective of whether it has a measure
to deliver or enough energy to transmit. Recall that such information
is indeed not available at the FC. Every $n$th IR is thus composed
by one frame with $M$ slots and has fixed duration $T_{IR}^{TD}=MT_{s}$,
as shown in Fig. \ref{fig:protocol_description}. Since TDMA is free
of communication errors in the considered interference-limited scenario,
its delivery probability $p_{d}(n)$ is only limited by energy availability
and it is thus an upper bound for ALOHA-based MACs. However, TDMA
might not be time efficient due to the many empty slots when the probability
of having a new measure $\alpha$ and/or the EH rate are small.

\subsubsection{Framed-ALOHA (FA) and Dynamic-FA (DFA)\label{Sec: FramedAloha_Description}}

Hereafter we describe the DFA protocol only, since FA follows as a
special case of DFA with no retransmissions capabilities as discussed
below. The $n$th IR, of duration $T_{IR}^{DFA}(n)$, is organized
into a set of frames as shown in Fig. \ref{fig:protocol_description}.
The \emph{backlog} $\mathcal{B}_{k}(n)$ for the $k$th frame is the
set composed of all sensors that simultaneously satisfy the following
three conditions: \textit{i}) have a new measure to transmit in the
$n$th IR; \textit{ii}) have transmitted unsuccessfully (because of
collisions) in the previous $k-1$ frames (this condition does not
apply for frame $k=1$); \textit{iii}) have enough energy left in
the ESD to transmit in the $k$th frame. All the users in the set
$\mathcal{B}_{k}(n)$, whose cardinality $\left\vert \mathcal{B}_{k}(n)\right\vert =B_{k}(n)$\textit{\emph{
is referred to as }}\textit{backlog size}, thus attempt transmission
during frame $k$. To make this possible, the FC allocates a frame
of $L_{k}(n)$ slots, where $L_{k}(n)$ is selected based on the estimate
$\hat{B}_{k}(n)$ of the backlog size $B_{k}(n)$ (estimation of $B_{k}(n)$
is discussed in Sec. \ref{Sec: BacklogEstimation}) as 
\begin{equation}
L_{k}(n)=\left\lceil \rho\hat{B}_{k}(n)\right\rceil ,\label{eqn: FrameSizeFA}
\end{equation}
where $\left\lceil \cdot\right\rceil $ is the upper nearest integer
operator, and $\rho$ is a design parameter. Note that, if the backlog
size is $B$, the probability $\beta\left(j,B,L\right)$ that $j\leq B$
sensors transmit in the same slot in a frame of length $L$ is binomial
\cite{art: Lucent} 
\begin{equation}
\beta\left(j,B,L\right)=\binom{B}{j}\left(\frac{1}{L}\right)^{j}\left(1-\frac{1}{L}\right)^{B-j}.\label{eqn: Slot_Occupation_Probabilities2_binomial}
\end{equation}
 Finally, FA is a special case of DFA where only one single frame
of size $L_{1}(n)$ is announced as retransmissions are not allowed
within the same IR.

\section{Analysis of the MAC Performance Metrics\label{Sec: Analysis_Performance_Metrics}}

In this section we derive the performance metrics defined in Sec.
\ref{Sec: MACperformanceMetrics} for TDMA, FA and DFA. The analysis
is based on two simplifying assumptions: 
\begin{itemize}
\item [$\mathcal{A}.1$] \textit{Known backlog}: the FC knows the backlog
size $B_{k}(n)=\left|\mathcal{B}_{k}(n)\right|$ before each $k$th
frame; 
\item [$\mathcal{A}.2$] \textit{Large backlog}: the backlog size $B_{k}(n)$,
in any IR $n$ and any frame $k$ of size $L_{k}(n)=\left\lceil \rho B_{k}(n)\right\rceil $,
is large enough to let the probability (\ref{eqn: Slot_Occupation_Probabilities2_binomial})
be approximated by the Poisson distribution \cite{art: Lucent}: 
\end{itemize}
\begin{equation}
\beta\left(j,B_{k}(n),L_{k}(n)\right)\simeq\frac{e^{-\frac{1}{\rho}}}{\rho^{j}j!}.\label{eqn: Poisson_Approximation}
\end{equation}

Assumption $\mathcal{A}.1$ simplifies the analysis as in reality
the backlog can only be estimated by the FC (see Sec. \ref{Sec: BacklogEstimation}
and Sec. \ref{Sec: Numerical_Results} for the impact of backlog estimation).
Assumption $\mathcal{A}.2$ is standard and analytically convenient,
as it makes the probability $\beta\left(j,B_{k}(n),L_{k}(n)\right)$
dependent only on the ratio $\rho$ between the frame length $L_{k}(n)$
and the backlog size $B_{k}(n)$. The assumptions above are validated
numerically in Sec. \ref{Sec: Numerical_Results}.

\subsection{Delivery Probabilities\label{Sec: Delivery_probabilities}}

Here we derive the delivery probability (\ref{eqn: DeliveryProbability})
within any $n$th IR under the assumptions $\mathcal{A}.1$ and $\mathcal{A}.2$
for the considered MAC protocols. The IR index $n$ is dropped to
simplify the notation.

\subsubsection{Delivery Probability for TDMA\label{Sec: Delivery_prob_TDMA}}

As the TDMA protocol is free of collisions, each sensor $S_{m}$ that
has a new measure to report in the current IR cannot deliver its payload
to the FC only when it is in energy shortage, namely if $E_{m}<\varepsilon$.
Provided that user $S_{m}$ has a new measure to transmit, the delivery
probability (\ref{eqn: DeliveryProbability}) reduces to 
\begin{equation}
p_{d}^{TD}=\Pr\left[E_{m}\geq\varepsilon\right]=G_{E}^{TD}\left(\varepsilon\right),\label{eqn: DeliveryProbTDMA}
\end{equation}
which is independent of the sensor index $m$ and dependent only on
the ccdf $G_{E}^{TD}\left(\cdot\right)$ of the energy stored in sensor
ESD at the beginning of the considered IR. The ESD energy distribution
for any arbitrary $n$th IR is derived in Sec. \ref{Sec: ESDenergyEvolution}.

\subsubsection{Delivery Probability for FA\label{Sec: Delivery_probability_FA}}

In the FA protocol, each sensor $S_{m}$ that has a new measure to
report in the current IR is able to correctly deliver its payload
to the FC only if: a) it transmits successfully in the selected slot,
possibly in the presence of interfering users provided that its SIR
is $\gamma_{m,1}\geq\gamma_{th}$; and b) it has enough energy to
transmit. From (\ref{eqn: Istantaneous_SIR}), the probability that
sensor $S_{m}$, with $S_{m}\in\mathcal{B}_{1}$, transmits successfully
in the selected slot, given that $\left\vert \mathcal{I}_{m,1}\right\vert =j$
users select the same slot of $S_{m}$ (thus colliding), is given
by 
\begin{equation}
p_{c}(j)=\Pr\left[h_{m}\geq\gamma_{th}\sum_{l=1}^{j}h_{l}\right],\label{eqn: P_c_j_FA}
\end{equation}
where, without loss of generality, we assumed that $\mathcal{I}_{m,1}=\{S_{1},...,S_{j}\}$,
and $S_{m}\notin\mathcal{I}_{m,1}$, as users are stochastically equivalent.
Under the large backlog assumption $\mathcal{A}.2$, the probability
that there are $j$ interfering users is Poisson-distributed (see
(\ref{eqn: Poisson_Approximation})), and thus the unconditional probability
$p_{c}$ that $S_{m}$ captures the selected slot can be approximated
as 
\begin{equation}
p_{c}\simeq e^{-\frac{1}{\rho}}\sum_{j=0}^{\infty}\frac{1}{\rho^{j}j!}p_{c}(j).\label{eqn: Capture_prob_FA}
\end{equation}
Note that, in (\ref{eqn: Capture_prob_FA}) we also extended the number
of possible interfering users up to infinity as $p_{c}(j)$ rapidly
vanishes for increasing $j$. Moreover, depending on the channel gain
pdf $f_{h}(\cdot)$, probabilities (\ref{eqn: P_c_j_FA}) can be calculated
either analytically (e.g., when $f_{h}(\cdot)$ is exponential, see
\cite{art: boyd}) or numerically.

Finally, under assumption $\mathcal{A}.2$, the successful transmission
event is independent of the ESD energy levels (which in principle
determine the actual backlog size in (\ref{eqn: Slot_Occupation_Probabilities2_binomial})),
and thus the delivery probability (\ref{eqn: DeliveryProbability})
for the FA protocol can be calculated as the product between the probability
$G_{E}^{FA}\left(\varepsilon\right)=\Pr\left[E_{m}\geq\varepsilon\right]$
that sensor $S_{m}$ has enough energy to transmit and the (approximated)
capture probability (\ref{eqn: Capture_prob_FA}) as 
\begin{equation}
p_{d}^{FA}\simeq G_{E}^{FA}\left(\varepsilon\right)e^{-\frac{1}{\rho}}\sum_{j=0}^{\infty}\frac{1}{\rho^{j}j!}p_{c}(j),\label{eqn: Prob_Delivery_FA}
\end{equation}
where the ESD energy ccdf $G_{E}^{FA}\left(\varepsilon\right)$ for
any arbitrary $n$th IR is derived in Sec. \ref{Sec: ESDenergyEvolution}.

\subsubsection{Delivery Probability for DFA\label{Sec: ProbDeliveryDFA}}

DFA is composed of several instances of FA, one for each $k$th frame
of the current IR. As DFA allows retransmissions, we need to calculate
the probability $p_{c,k}(j)$ that any sensor active during frame
$k$, say $S_{m}\in\mathcal{B}_{k}$, transmits successfully in the
selected slot given that there are $\left\vert \mathcal{I}_{m,k}\right\vert =j$
users that transmit in the same slot, with $\mathcal{I}_{m,k}\subseteq\mathcal{B}_{k}$.
The computation of $p_{c,k}(j)$, for $k>1$, is more involved than
(\ref{eqn: P_c_j_FA}). In fact, packets collisions introduce correlation
among the channel gains of collided users, as any sensor in the backlog
$\mathcal{B}_{k}$, for $k>1$, might have collided with some other
sensors in the set $\mathcal{B}_{k}$. We recall that, even though
the channel gains are i.i.d. at the beginning of the IR, they remain
fixed for the entire IR.

While the exact computation of probabilities $p_{c,k}(j)$ is generally
cumbersome, the large backlog assumption $\mathcal{A}.2$ enables
some simplifications. Specifically, correlation among channel gains
can be neglected, since for large backlogs it is unlikely that two
users collide more than once within the same IR. By assuming independence
among the channel gains at any frame, calculation of $p_{c,k}(j)$
requires only to evaluate the channel gain pdf $f_{h}^{(k)}(\cdot)$
at the $k$th frame for any user within $\mathcal{B}_{k}$, which
is the same for all users by symmetry. The computation of pdf $f_{h}^{(k)}(\cdot)$
can be done recursively, starting from frame $k=1$, so that at frame
$k$ we condition on the event that the SIR (\ref{eqn: Istantaneous_SIR})
was $\gamma_{m,k-1}<\gamma_{th}$. Under assumption $\mathcal{A}.2$,
this can be done numerically.

Now, let $\tilde{h}_{m}^{(k)}$, for $m\in\{1,...,M\}$ and $k\in\{1,...,F_{\varepsilon}\}$,
be random variables with pdf $f_{h}^{(k)}(\cdot)$ independent over
$m$, where $\tilde{h}_{m}^{(1)}=h_{m}$. The conditional capture
probabilities $p_{c,k}(j)$ can then be approximated as (compare to
(\ref{eqn: P_c_j_FA})) 
\begin{equation}
p_{c,k}(j)\simeq\Pr\left[\tilde{h}_{m}^{(k)}\geq\gamma_{th}\sum_{l=1}^{j}\tilde{h}_{l}^{(k)}\right],\label{eqn: p_c_k_j_DFA}
\end{equation}
for any $m\notin\{1,...,j\}$ as users are stochastically equivalent.
By exploiting the Poisson approximation similarly to (\ref{eqn: Capture_prob_FA}),
the unconditional probability that any user within the backlog successfully
transmits in the selected slot during the $k$th frame becomes 
\begin{equation}
p_{c,k}\simeq e^{-\frac{1}{\rho}}\sum_{j=0}^{\infty}\frac{1}{\rho^{j}j!}p_{c,k}(j).\label{eqn: p_c_k_DFA}
\end{equation}

Recalling that a user keeps retransmitting its message until it is
successfully delivered to the FC, then the successful delivery of
a message in a frame is a mutually exclusive event with respect to
the delivery in previous frames. Therefore, the probability of transmitting
successfully in the $k$th frame, given that enough energy is available,
is $p_{c,k}\prod_{i=1}^{k-1}\left(1-p_{c,i}\right).$ Finally, by
accounting for the probability $G_{E}^{DFA}\left(k\varepsilon\right)=\Pr\left[E_{m}\geq k\varepsilon\right]$
of having enough energy in each $k$th frame, the DFA delivery probability
can be obtained, under assumption $\mathcal{A}.2$, as%
\footnote{\label{foot: correlated_backlogs}Note that in principle the backlogs
$\mathcal{B}_{1},\mathcal{B}_{2}...$ are correlated, and therefore
the exact $p_{d}^{DFA}$ should be obtained by averaging over the
joint distribution of the backlog sizes. However, the assumption $\mathcal{A}.2$
removes the dependence on the backlog size.%
} 
\begin{equation}
p_{d}^{DFA}\simeq\sum_{k=1}^{F_{\varepsilon}}G_{E}^{DFA}\left(k\varepsilon\right)p_{c,k}\prod_{i=1}^{k-1}\left(1-p_{c,i}\right),\label{eqn: Delivery_Probability_DFA_Whole2}
\end{equation}
where the ESD energy ccdf $G_{E}^{DFA}\left(k\varepsilon\right)$
for any arbitrary $n$th IR is derived in Sec. \ref{Sec: ESDenergyEvolution}.

\subsection{Time Efficiencies\label{Sec: Time_efficiency}}

In this section we derive the time efficiency (\ref{eqn: Time_efficiency})
for the three considered protocols.

\subsubsection{Time Efficiency for TDMA}

Let $\mathcal{M}_{m}$ be the event indicating that user $S_{m}$
has a new measure to report in the current IR, with $\Pr[\mathcal{M}_{m}]=\alpha$,
then the TDMA time efficiency (\ref{eqn: Time_efficiency}) is given
by the probability that the $m$th user has enough energy to transmit
and a new measure to report: 
\begin{equation}
p_{t}^{TD}=\Pr\left[E_{m}\geq\varepsilon,\mathcal{M}_{m}\right]=\Pr\left[E_{m}\geq\varepsilon\right]\Pr\left[\mathcal{M}_{m}\right]=\alpha G_{E}^{TD}\left(\varepsilon\right),\label{eqn: Time_Efficiency_TDMA}
\end{equation}
where we exploited independence between energy availability $E_{m}$
and $\mathcal{M}_{m}$.

\subsubsection{Time Efficiency for FA\label{Sec: TimeEfficiencyFA}}

Since we assumed $\gamma_{th}>0dB$, then when more than one user
transmits within the same slot, only one of them can be decoded successfully,
that is, successful transmissions of different users within the same
slot are disjoint events. Therefore, the probability that a slot,
simultaneously selected by $j$ users, is successfully used by any
of them is given by $jp_{c}(j-1)$, where $p_{c}(j-1)$ is (\ref{eqn: P_c_j_FA})
by recalling that any user have $(j-1)$ interfering users. Furthermore,
under assumption $\mathcal{A}.2$, the probability that exactly $j$
users select the same slot is $e^{-\frac{1}{\rho}}/\left(\rho^{j}j!\right)$,
and by summing up over the number of simultaneously transmitting users
$j$ we get
\begin{equation}
p_{t}^{FA}\simeq e^{-\frac{1}{\rho}}\sum_{j=1}^{\infty}\frac{1}{\rho^{j}j!}jp_{c}(j-1)=e^{-\frac{1}{\rho}}\sum_{j=0}^{\infty}\frac{1}{\rho^{(j+1)}j!}p_{c}\left(j\right)\label{eqn: Time_Efficiency__FA}
\end{equation}
Note that, a consequence of assumption $\mathcal{A}.2$ is to make
the FA time efficiency (\ref{eqn: Time_Efficiency__FA}) independent
of the ESD energy distribution. Moreover we remark that, when $\rho=1$,
$p_{c}(j)=1$ for $j=0$ and $p_{c}(j)=0$ for $j>0$, then we have
$p_{t}^{FA}=e^{-1}$, which is the throughput of slotted ALOHA \cite{art: Schoute}.

\subsubsection{Time Efficiency for DFA\label{Sec: TimeEfficiencyDFA}}

The derivation of the DFA time efficiency $p_{t}^{DFA}$ follows from
the FA time efficiency by accounting for the presence of multiple
frames within an IR similarly to Sec. \ref{Sec: ProbDeliveryDFA}.
Since the time efficiency is defined over multiple frames, we first
derive the time efficiency in the $k$th frame, similarly to (\ref{eqn: Time_Efficiency__FA})
but considering (\ref{eqn: p_c_k_j_DFA}) instead of (\ref{eqn: P_c_j_FA}),
as 
\begin{equation}
p_{t,k}^{DFA}\simeq e^{-\frac{1}{\rho}}\sum_{j=0}^{\infty}\frac{1}{\rho^{(j+1)}j!}p_{c,k}\left(j\right).\label{eqn: Time_efficiency_DFA_conditional}
\end{equation}
We then calculate $p_{t}^{DFA}$ by summing (\ref{eqn: Time_efficiency_DFA_conditional})
up, for all $k\in\{1,...,F_{\varepsilon}\}$, weighted by the (random)
length of the corresponding frame $L_{k}$ normalized to the total
number of slots in the IR $\sum_{k=1}^{F_{\varepsilon}}L_{k}$. Note
that, under assumption $\mathcal{A}.2$ the random frame length $L_{k}$
is well-represented by its (deterministic) average value $L_{k}\simeq E\left[L_{k}\right]=\rho E\left[B_{k}\right]$
and thus the DFA time efficiency results 
\begin{equation}
p_{t}^{DFA}\simeq\frac{\sum_{k=1}^{F_{\varepsilon}}p_{t,k}^{DFA}E\left[B_{k}\right]}{\sum_{k=1}^{F_{\varepsilon}}E\left[B_{k}\right]},\label{eqn: Time_efficiency_DFA}
\end{equation}
where the average backlog size $E[B_{k}]$ in frame $k$, can be computed,
under assumption $\mathcal{A}.2$, as $E[B_{k}]=M\alpha G_{E}^{DFA}(k\varepsilon)\prod_{i=1}^{k-1}\left(1-p_{c,i}\right)$.
In fact, $M\alpha$ indicates the average number of users that have
a new measure to report in the current IR, $G(k\varepsilon)$ is the
probability that $k\varepsilon$ energy units are stored in the ESD
at the beginning of the IR, thus allowing $k$ successful transmissions,
and $\prod_{i=1}^{k-1}\left(1-p_{c,i}\right)$ is the probability
that a sensor collides in all of the first $(k-1)$ frames.

\section{ESD energy evolution\label{Sec: ESDenergyEvolution}}

In Sec. \ref{Sec: Analysis_Performance_Metrics} we have shown that
the performance metrics for the $n$th IR depend on the energy distribution
in the sensor ESD at the beginning of the IR itself. The goal of this
section is to derive the ccdf $G_{E(n)}(\cdot)$, for any IR $n$,
in order to obtain the asymptotic performance metrics (\ref{eqn: Asymptotic_Delivery_Probability})
and (\ref{eqn: Asymptotic_Time_efficiency}) from Sec. \ref{Sec: Delivery_probabilities}
and Sec. \ref{Sec: Time_efficiency} respectively.

\begin{figure}[h!]
\centering \includegraphics[clip,width=10cm]{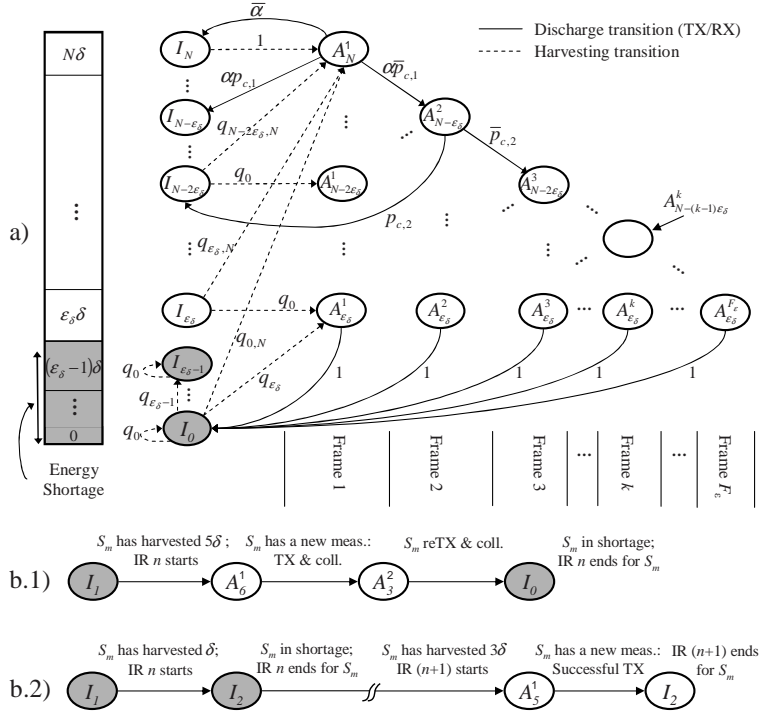}
\caption{a) Discrete Markov chain used to model the evolution of the energy
stored in the discrete ESD of a sensor in terms of the energy unit
$\delta$. In b.1) and b.2) there are two outcomes of possible state
transition chains for $\varepsilon_{\delta}=3$. Grey shaded states
indicate energy shortage condition. Some transitions are not depicted
to simplify representation. ($\bar{\alpha}=1-\alpha$ and $\bar{p}_{c,k}=1-p_{c,k}$).}

\label{fig:DMC_model}
\end{figure}

In general, the evolution of sensor ESDs across IRs in DFA are correlated
with each other, due to the possibility of retransmitting after collisions.
However, under the large backlog assumption $\mathcal{A}.2$, similarly
to the discussion in Sec. \ref{Sec: ProbDeliveryDFA}, the evolution
of sensor ESDs become decoupled and can thus be studied separately.
Accordingly, we develop a stochastic model, based on a discrete Markov
chain (DMC) that focuses on a single sensor ESD as shown in Fig. \ref{fig:DMC_model}.
In addition, we concentrate on the DFA protocol as ESD evolutions
for TDMA and FA follow as special cases. Note that, in TDMA (or FA),
the evolution of sensor ESDs are actually independent with each other
as retransmissions are not required (or allowed).

\subsection{States of a Sensor\label{Sec: EnergyDiscretization}}

The state of a sensor is uniquely characterized by: \textit{i}) sensor
activity or idleness (see below); \textit{ii}) the amount of energy
stored in its ESD; \textit{iii}) the current frame index if the sensor
is active. A sensor is \textit{active} if it has a new measure still
to be delivered to the FC in the current IR and enough energy in its
ESD, while it is \textit{idle} otherwise. States in which the sensor
is active, referred to as \textit{active states}, are denoted by $A_{j}^{k}$
and they are characterized by: \textit{a}) the current frame index
$k\in\{1,...,F_{\varepsilon}\}$; and \textit{b}) the number $j\in\{0,...,N\}$
of energy units $\delta$ stored in the sensor ESD.

States in which the sensor is idle, referred to as \textit{idle states},
are instead denoted by $I_{j}$ and they are uniquely characterized
by the number $j\in\{0,...,N\}$ of energy units stored in the sensor
ESD. EH is then associated to idle states given the assumption that
any energy arrival in the current IR can only be used in the next
IR (see Sec. \ref{Sec: Energy_Harvesting_model}).

\subsection{Discrete Markov Chain (DMC) Model\label{Sec: DMCmodel}}

\begin{figure}[h!]
\centering \includegraphics[clip,width=10cm]{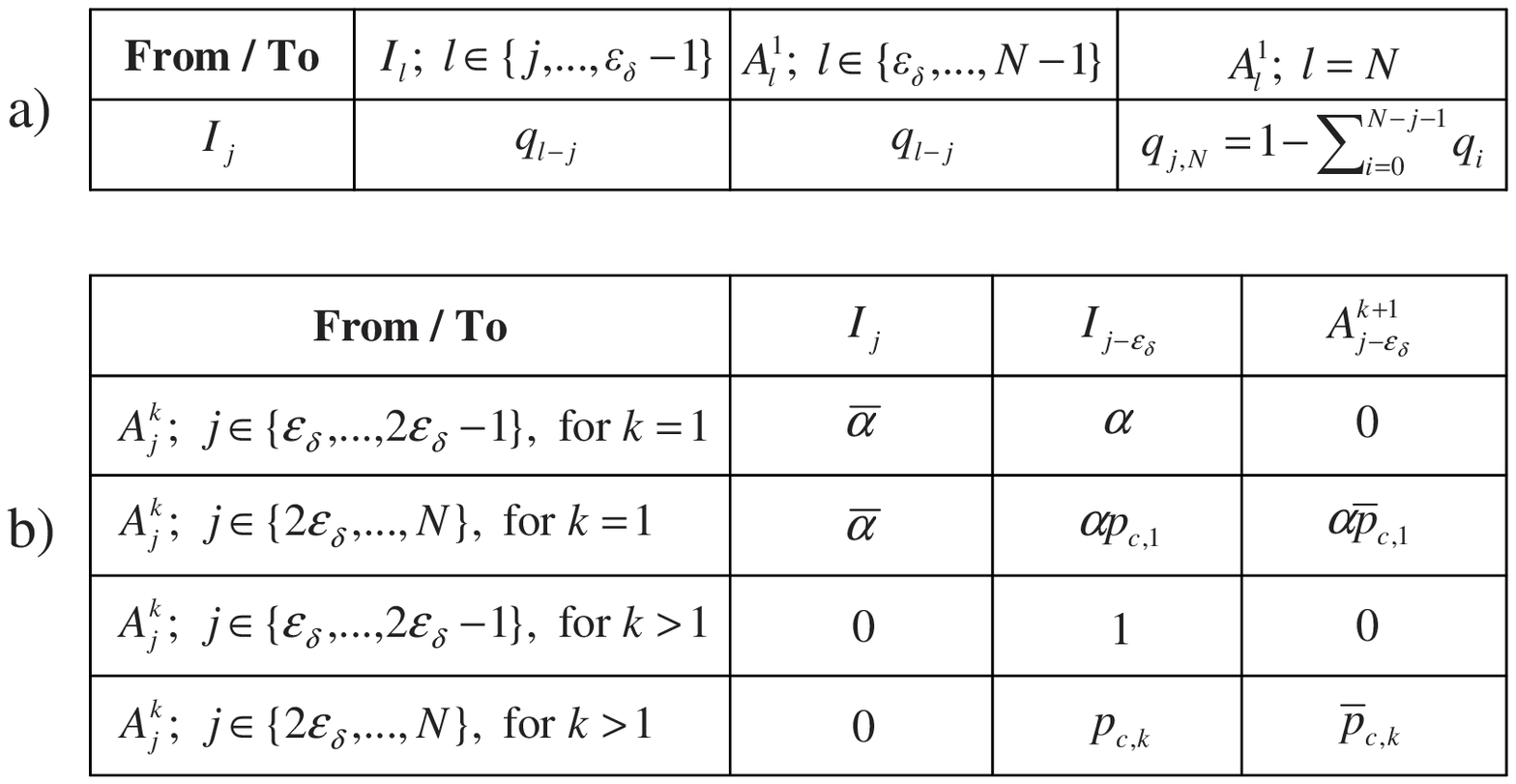}
\caption{State transition probabilities for the DMC model in Sec. \ref{Sec: DMCmodel}:
a) transition probabilities due to energy harvesting; b) transition
probabilities due to the bidirectional communication with the FC.
The transition matrix $\mathbf{P}$ can be derived according to the
probabilities in a) and b) for all the values of $k\in\{1,...,F_{\varepsilon}\}$
and $j\in\{0,...,N\}$.}

\label{fig:transition_table}
\end{figure}

Operations of a sensor across IRs are as follows. When sensor $S_{m}$
is not involved in an IR, it is in an idle state, say $I_{j}$, waiting
for the next IR. When a new IR begins, the energy harvested in the
last interval $T_{int}$ is added, so that, if the ESD is not in energy
shortage, the state makes a transition $I_{j}$ $\rightarrow A_{l}^{1}$
toward an active state, with $l\geq\varepsilon_{\mathcal{\delta}}\geq j$.
Otherwise, if it is in energy shortage, it makes a transition $I_{j}\rightarrow I_{l}$
toward another idle state, with $j\leq l<\varepsilon_{\mathcal{\delta}}$.
If sensor $S_{m}$ is not in energy shortage, it remains in state
$A_{j}^{1}$ at the beginning of the IR only if it has a new measure
to transmit, which happens with probability $\alpha$. Instead, with
probability $\bar{\alpha}=1-\alpha$ the state makes a transition
toward an idle state as $A_{j}^{1}\rightarrow I_{j}$. If there is
a new measure, the sensor keeps transmitting it in successive frames
until either the packet is correctly delivered to the FC, or its ESD
falls in energy shortage, or both. A collision in frame $k$ happens
with probability $\bar{p}_{c,k}=1-p_{c,k}$ (see Sec. \ref{Sec: ProbDeliveryDFA})
and leads to a transition either $A_{j}^{k}\rightarrow A_{j-\varepsilon_{\mathcal{\delta}}}^{k+1}$,
for $j\geq2\varepsilon_{\mathcal{\delta}}$ (no shortage after collision)
or $A_{j}^{k}\rightarrow I_{j-\varepsilon_{\mathcal{\delta}}}$, for
$j<2\varepsilon_{\mathcal{\delta}}$ (shortage after collision). Successful
transmissions in frame $k$, which happens with probability $p_{c,k}$,
instead leads to a transition $A_{j}^{k}\rightarrow I_{j-\varepsilon_{\mathcal{\delta}}}$.
Transition probabilities are summarized in Fig. \ref{fig:transition_table},
where we have defined $q_{j,N}=\Pr[E_{H,m}\geq(N-j)\mathcal{\delta}]=1-\sum_{i=0}^{N-j-1}q_{i}$.
Note that, the probability $\alpha$ of having a new measure is only
accounted for in active states in the first frame (i.e., in states
$A_{j}^{1}$, for $j\in\{0,...,N\}$, see Fig. \ref{fig:transition_table}-b)).
In fact, being in any state $A_{j}^{k}$ for $k>1$ already implies
that a new measure was available at the beginning of the IR. Notice
that, according to the model above, state transitions in the DMC at
hand are event-driven and do not happen at fixed time intervals. A
sketch of the considered DMC is shown in Fig. \ref{fig:DMC_model}-a),
while we show two outcomes of possible state transition chains in
Fig. \ref{fig:DMC_model}-b.1) and \ref{fig:DMC_model}-b.2).

From Fig. \ref{fig:DMC_model}-a), it can be seen that, when $q_{0}>0$,
$q_{1}>0$ and $p_{c,k}>0$, for $k\in\{1,...,F_{\varepsilon}\}$,
the DMC at hand is irreducible and aperiodic and thus, by definition,
ergodic (see \cite{lib: Gallager}). In fact, if $q_{1}>0$, any state
of the Markov model can be reached from any other state with non-zero
probability, and therefore the Markov chain is irreducible. Moreover,
the probability of having a self-transition from state $I_{0}$ to
itself is $q_{0}>0,$ and therefore state $I_{0}$ is aperiodic. The
presence of an aperiodic state in a finite state irreducible Markov
chain is enough to conclude that the chain is aperiodic \cite[Ch. 4, Th. 1]{lib: Gallager}.
Since the DMC is ergodic it admits a unique steady-state probability
distribution $\mathbf{\phi=[}\phi_{I_{0}},...,\phi_{I_{N}},\phi_{A_{\varepsilon_{\mathcal{\delta}}}^{1}},...,\phi_{A_{N}^{F_{\varepsilon}}}]$,
regardless of the initial distribution, which can be calculated by
resorting to conventional techniques \cite{lib: Gallager}. This also
guarantees the existence of limits (\ref{eqn: Asymptotic_Delivery_Probability})
and (\ref{eqn: Asymptotic_Time_efficiency}). Vector $\mathbf{\phi}$
represents the steady-state distribution in any discrete time instant
of the interrogation period (i.e., during either any frames of an
IR or idle period). However, to calculate (\ref{eqn: Asymptotic_Delivery_Probability})
and (\ref{eqn: Asymptotic_Time_efficiency}) we need the DMC steady-state
distribution $\mathbf{\phi}^{+}$ conditioned on being at the beginning
of the IR. This can be calculated by recalling that between the end
of the last issued IR and the beginning of a new one, sensor $S_{m}$
can only be in any idle states $I_{j}$, with $j\in\{0,...,N\}$,
and thus its state conditional distribution $\mathbf{\phi}^{-}\mathbf{=[}\phi_{I_{0}}^{-},...,\phi_{I_{N}}^{-},\phi_{A_{\varepsilon_{\mathcal{\delta}}}^{1}}^{-},...,\phi_{A_{N}^{F_{\varepsilon}}}^{-}]$,
is given by $\phi_{I_{j}}^{-}=\phi_{I_{j}}/\sum_{i=0}^{N}\phi_{I_{i}}$,
$\forall j\in\{0,...,N\}$ and $\phi_{A_{j}^{k}}^{-}=0$, for all
$j,k$. The desired distribution $\mathbf{\phi}^{+}$ of the state
at the beginning of the next IR can be obtained as $\mathbf{\phi}^{+}=\mathbf{\phi}^{-}\mathbf{P}$,
where $\mathbf{P}$ is the DMC probability transition matrix of the
DMC in Fig. \ref{fig:DMC_model}-a) that can be obtained through Fig.
\ref{fig:transition_table}. Note that, according to the transition
probabilities in Fig. \ref{fig:transition_table}, starting from any
state $I_{j}$, with $j\in\{0,...,N\}$, only states $I_{j}$, with
$j\in\{0,...,\varepsilon_{\delta}-1\}$ and states $A_{j}^{1}$, with
$j\in\{\varepsilon_{\delta},...,N\}$ can be reached. Therefore, the
only possibly non-zero entries of distribution $\mathbf{\phi}^{+}$
are $\phi_{I_{j}}^{+}\text{ for }j\in\{0,...,\varepsilon_{\delta}-1\}$
and $\phi_{A_{j}^{1}}^{+}\text{for }j\in\{\varepsilon_{\delta},...,N\}$.

Once the DMC steady-state distribution $\mathbf{\phi}^{+}$ at the
beginning of any (steady-state) IR is obtained, we can calculate the
steady-state distribution $p_{E(n\rightarrow\infty)}(\cdot)$ of the
energy stored in the sensor ESD at the beginning of any (steady-state)
IR, denoted by $\mathbf{\pi}_{E}=[\pi_{E}(0),...,\pi_{E}(N)]$, by
mapping the DMC states into the energy level set $\mathcal{E}$ as
follows 
\begin{equation}
\pi_{E}(j)=\left\{ \begin{array}{ll}
\phi_{I_{j}}^{+} & \text{ for }j\in\{0,...,\varepsilon_{\delta}-1\}\\
\phi_{A_{j}^{1}}^{+} & \text{for }j\in\{\varepsilon_{\delta},...,N\}
\end{array}\right..\label{eqn: Steady_state_ESD_Energy}
\end{equation}
The ccdf $G_{E(n\rightarrow\infty)}(\cdot)$ is immediately derived
from $\mathbf{\pi}_{E}$. Finally, we remark that analysis of FA and
TDMA can be obtained by limiting the set of active states to $A_{\varepsilon_{\delta}}^{1},A_{\varepsilon_{\delta}+1}^{1},...,A_{N}^{1}$
(i.e., no retransmission), and recalling that sensor $S_{m}$ after
transmission returns to idle states regardless of the success of transmission.\newpage

\section{Backlog Estimation\label{Sec: BacklogEstimation}}

In this section we propose a backlog estimation algorithm for the
DFA protocol (extension to FA is straightforward). Unlike previous
work on the subject \cite{art: Lucent}\cite{art: MLbacklogEstim},
here backlog estimation is designed by accounting for the interplay
of EH, capture effect and multiple access. Computational complexity
of optimal estimators is generally intractable for a large number
of sensors even for conventional systems (see e.g., \cite{art: MLbacklogEstim}).
We thus propose a low-complexity two-steps backlog estimation algorithm
that, neglecting the IR index, operates in every IR as follows: \textit{i})
the FC estimates the initial backlog size $B_{1}$ based on the ccdf
$G_{E}\left(\varepsilon\right)$ of the ESD energy at the beginning
of the current IR; \textit{ii}) the backlog estimates for the next
frames are updated based on the channel outcomes and the residual
ESD energy.

For the first frame, the backlog size estimate and the frame length
are $\hat{B}_{1}=M\alpha G_{E}\left(\varepsilon\right)$ and $L_{1}=\left\lceil \rho\hat{B}_{1}\right\rceil $,
respectively. For subsequent frames, let us assume that the FC announced
a frame of $L_{k}=\left\lceil \rho\hat{B}_{k}\right\rceil $ slots.
The FC estimates the backlog size for frame $k+1$ by counting the
number of slots that are successful ($N_{D,k}$) and collided ($N_{C,k}$)
within the $k$th frame of length $L_{k}$ slots. Since the FC cannot
discern exactly how many sensors transmitted in each successful slot,
the estimate of the total number $C_{D,k}$ of sensors that collided
in $N_{D,k}$ successful slots is $\hat{C}_{D,k}=\left(\beta_{D,k}-1\right)N_{D,k}$,
with $\beta_{D,k}$ being the conditional average number of sensors
that transmit in a slot given that the slot is successful (with no
capture $\beta_{D,k}=1$). Similarly, for the collided slots we obtain
$\hat{C}_{C,k}=\beta_{C,k}N_{C,k}$, where $\beta_{C,k}$ is now conditioned
on observing a collided slot. Derivations of $\beta_{D,k}$ and $\beta_{C,k}$
are in Appendix \ref{App: AVGnumberOfSensors}. Since the estimate
of the total number of sensors that unsuccessfully transmitted is
$\hat{C}_{k}=\hat{C}_{C,k}+\hat{C}_{D,k}$, the backlog size estimate
$\hat{B}_{k+1}$ for the $(k+1)$th frame is obtained by accounting
for the fraction of sensors within $\hat{C}_{k}$ that are not in
energy shortage: $\hat{B}_{k+1}=\hat{C}_{k}G_{E}((k+1)\varepsilon|k\varepsilon)$,
where $G_{E}((k+1)\varepsilon|k\varepsilon)=\Pr\left[E_{m}\geq(k+1)\varepsilon|E_{m}\geq k\varepsilon\right]$.
The proposed backlog estimation scheme thus works as follows: 
\begin{equation}
\hat{B}_{k}=\left\{ \begin{array}{lc}
M\alpha G_{E}\left(\varepsilon\right) & \text{if }k=1\\
\hat{C}_{k-1}G_{E}(k\varepsilon|\left(k-1\right)\varepsilon) & \text{if }k>1
\end{array}\right..\label{eqn: BacklogEstimationGeneral}
\end{equation}
Algorithm (\ref{eqn: BacklogEstimationGeneral}) can be applied to
any $n$th IR by deriving the ESD distribution $p_{E(n)}(\cdot)$
(or $G_{E(n)}(\cdot)$) from any initial distribution $p_{E(1)}(\cdot)$,
by exploiting the DMC model in Sec. \ref{Sec: DMCmodel}.

\section{Numerical Results\label{Sec: Numerical_Results}}

In this section, we present extensive numerical results to get insight
into the MAC protocols design. Moreover, to validate the analysis
proposed in Sec. \ref{Sec: Analysis_Performance_Metrics} and Sec.
\ref{Sec: ESDenergyEvolution}, we compare the analytical results
therein with a simulated system that does not rely on simplifying
assumptions $\mathcal{A}.1$ and $\mathcal{A}.2$. The performances
of the backlog estimation algorithm proposed in Sec. \ref{Sec: BacklogEstimation}
are also assessed through a comparison with the ideal case of perfectly
known backlog at the FC.

\subsection{MAC Performance Metrics Trade-offs\label{Sec: Numerica_MACtradeoffs}}

The energy $E_{H,m}(n)$ harvested between two successive IRs is assumed
as geometrically-distributed with $q_{i}=\Pr[E_{H,m}(n)=i\delta]=\xi(1-\xi)^{i}$,
where $\xi=\delta/(\delta+\mu_{H})$. The average harvested energy
normalized by $\varepsilon$, referred to as \textit{harvesting rate},
is $E[E_{H,m}(n)/\varepsilon]=\mu_{H}$.

\begin{figure}[h!]
\centering \includegraphics[clip,width=10cm]{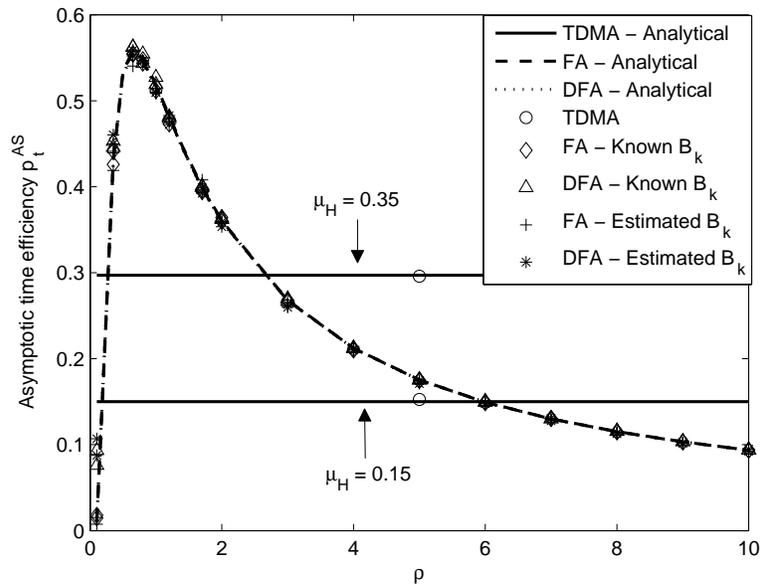} \caption{Asymptotic time efficiency (\ref{eqn: Asymptotic_Time_efficiency})
versus $\rho$, for different energy harvesting rates $\mu_{H}\in\{0.15,0.35\}$.
Comparisons are between analytical derivations and simulated results
with both known ($B_{k}$) and estimated backlog ($\hat{B}_{k}$,
see (\ref{eqn: BacklogEstimationGeneral})), ($M=400$, $\gamma_{th}=3dB$,
$\alpha=0.3$, $F_{\varepsilon}=10$, $\varepsilon=1$, $\delta=1/50$).}

\label{fig:time_efficiency}
\end{figure}

The asymptotic time efficiencies (\ref{eqn: Asymptotic_Time_efficiency})
for TDMA, FA and DFA protocols, are shown in Fig. \ref{fig:time_efficiency}
versus design parameter $\rho$ (recall (\ref{eqn: FrameSizeFA})).
System performance is evaluated by considering: $\mu_{H}\in\{0.15,0.35\}$,
$M=400$, $\gamma_{th}=3dB$, $\alpha=0.3$; $\varepsilon$ is normalized
to unity, energy unit is $\delta=1/50$ so that $\varepsilon_{\delta}=50$
and $F_{\varepsilon}=10$. We compare the analytical performance metrics
derived in Sec. \ref{Sec: Analysis_Performance_Metrics} with simulated
scenarios for both known and estimated backlog. While the performance
of TDMA is clearly independent of $\rho$, in FA and DFA there is
a time efficiency-maximizing $\rho$, which is close to one (in \cite{art: Schoute}
the optimal value was $\rho=1$ since the capture effect was not considered).
The effect of decreasing (or increasing) the harvesting rate $\mu_{H}$
on the TDMA time efficiency is due to the larger (or smaller) number
of sensors that are in energy shortage and whose slots are not used,
while it is negligible for FA and DFA due to their ability to dynamically
adjust the frame size according to backlog estimates $\hat{B}_{k}$.
The tight match between analytical and simulated results also validates
assumptions $\mathcal{A}.1$ and $\mathcal{A}.2$ and the efficacy
of the backlog estimation algorithm.

\begin{figure}[h!]
\centering \includegraphics[clip,width=10cm]{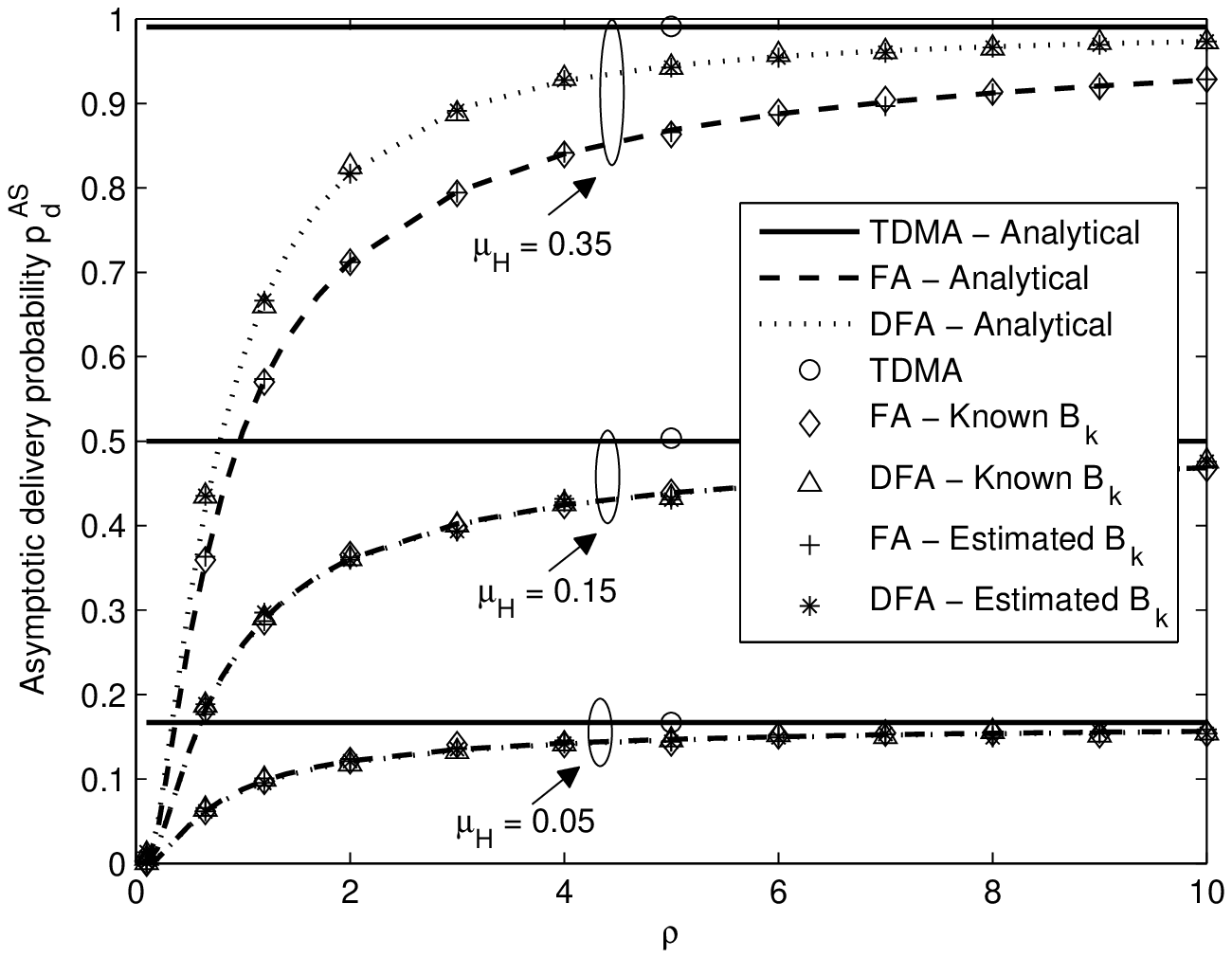}
\caption{Asymptotic delivery probability (\ref{eqn: Asymptotic_Delivery_Probability})
versus $\rho$, for different energy harvesting rate $\mu_{H}\in\{0.05,0.15,0.35\}$.
Comparisons are between analytical derivations and simulated results
with both known ($B_{k}$) and estimated backlog ($\hat{B}_{k}$,
see (\ref{eqn: BacklogEstimationGeneral})), ($M=400$, $\gamma_{th}=3dB$,
$\alpha=0.3$, $F_{\varepsilon}=10$, $\varepsilon=1$, $\delta=1/50$).}

\label{fig:delivery_probability}
\end{figure}

The asymptotic delivery probability, for harvesting rate $\mu_{H}$
$\in\{0.05,0.15,0.35\}$, versus design parameter $\rho$ is shown
in Fig. \ref{fig:delivery_probability} with the same system parameters
as for Fig. \ref{fig:time_efficiency}. Unlike for the time efficiency,
TDMA always outperforms FA and DFA in terms of delivery probability.
In fact, sensors operating with TDMA and FA have the same energy consumption
since they transmit at most once per IR, while possibly more than
once in DFA. However, TDMA does not suffer collisions and thus it
is able to eventually deliver more packets to the FC. The delivery
probability strongly depends on the harvesting rate $\mu_{H}$, which
influences the ESD energy distribution and consequently the energy
shortage probability. Moreover, DFA outperforms FA thanks to the retransmission
capability when the harvesting rate is relatively high (e.g., $\mu_{H}$$=0.35$),
while for low harvesting rate (e.g., $\mu_{H}$$\in\{0.05,0.15\}$)
DFA and FA perform similarly. In fact, for low harvesting rates, most
of the sensors are either in energy shortage or have very low energy
in their ESDs. Hence, most of the sensors that are not in energy shortage
are likely to have only one chance to transmit, and thus retransmission
opportunities provided by DFA are not leveraged.

\begin{figure}[h!]
\centering \includegraphics[clip,width=10cm]{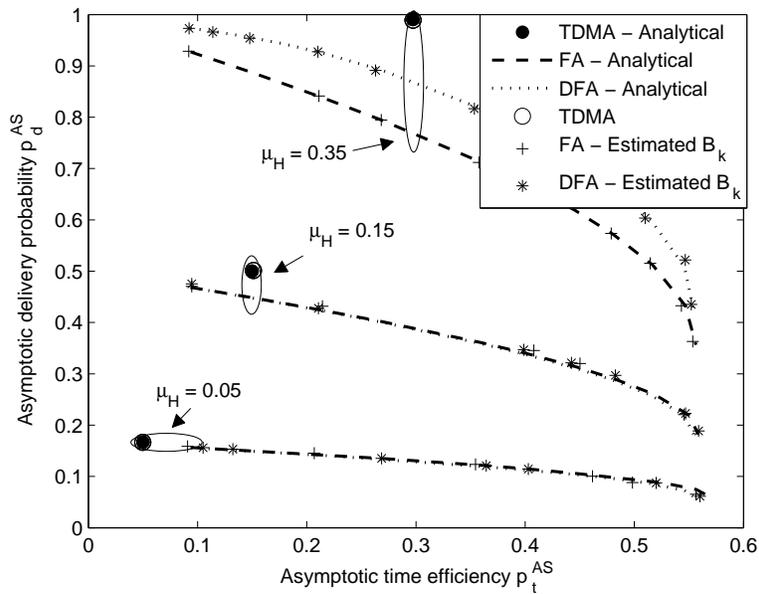}
\caption{Trade-off between asymptotic delivery probability (\ref{eqn: Asymptotic_Delivery_Probability})
and asymptotic time efficiency (\ref{eqn: Asymptotic_Time_efficiency})
for different energy harvesting rate $\mu_{H}\in\{0.05,0.15,0.35\}$.
Comparisons are between analytical derivations and simulated results
with estimated backlog ($\hat{B}_{k}$, see (\ref{eqn: BacklogEstimationGeneral})),
($M=400$, $\gamma_{th}=3dB$, $\alpha=0.3$, $F_{\varepsilon}=10$,
$\varepsilon=1$, $\delta=1/50$).}

\label{fig:efficiencies_region_fixed_gamma}
\end{figure}

The trade-off between asymptotic delivery probability (\ref{eqn: Asymptotic_Delivery_Probability})
and asymptotic time efficiency (\ref{eqn: Asymptotic_Time_efficiency})
is shown in Fig. \ref{fig:efficiencies_region_fixed_gamma} for different
values of the harvesting rate $\mu_{H}\in\{0.05,0.15,0.35\}$. System
parameters are the same as for Fig. \ref{fig:time_efficiency}. For
TDMA, the trade-off consists of a single point on the plane, whereas
FA and DFA allow for more flexibility via the selection of parameter
$\rho$. When increasing $\rho$ more sensors might eventually report
their measures to the FC, thus increasing the delivery probability
to the cost of lowering time efficiency (see Fig. \ref{fig:time_efficiency}
and \ref{fig:delivery_probability}). For FA and DFA, the trade-off
curves are obtained as $\max_{\rho}\left\{ p_{d}^{AS}\right\} $,
s.t. $p_{t}^{AS}=\lambda$ for each achievable $\lambda$.

\begin{figure}[h!]
\centering \includegraphics[clip,width=10cm]{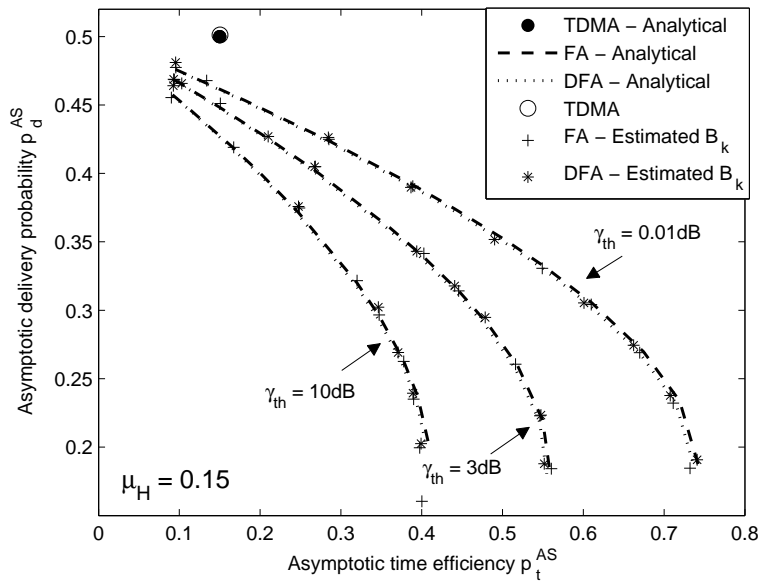}
\caption{Trade-off between asymptotic delivery probability (\ref{eqn: Asymptotic_Delivery_Probability})
and asymptotic time efficiency (\ref{eqn: Asymptotic_Time_efficiency})
for different SIR threshold $\gamma_{th}\in\{0.01,3,10\}dB$ values
and fixed energy harvesting rate $\mu_{H}=0.15$. Comparisons are
between analytical derivations and simulated results with estimated
backlog ($\hat{B}_{k}$, see (\ref{eqn: BacklogEstimationGeneral})),
($M=400$, $\alpha=0.3$, $F_{\varepsilon}=10$, $\varepsilon=1$,
$\delta=1/50$).}

\label{fig:efficiencies_region_var_gamma}
\end{figure}

The impact of the capture effect on the performance metrics trade-offs
is shown in Fig. \ref{fig:efficiencies_region_var_gamma}, where we
vary the SIR threshold $\gamma_{th}\in\{0.01,3,10\}dB$ and keep the
harvesting rate $\mu_{H}=0.15$ fixed (other parameters are as in
Fig. \ref{eqn: Asymptotic_Time_efficiency}). As expected, the lower
the SIR threshold $\gamma_{th}$ the higher the probability that the
SIR of any of the colliding sensors is above $\gamma_{th}$, and thus
the higher the performance obtained with ALOHA-based protocols. TDMA
is insensitive to $\gamma_{th}$.\newpage

\section{Conclusions}

The design of medium access control (MAC) protocols for single-hop
wireless sensor networks (WSNs) with energy-harvesting (EH) devices
offers new challenges as compared to the standard scenario with battery-powered
(BP) nodes. New performance criteria are called for, along with new
design solutions. This paper addresses these issues by investigating
the novel trade-off between the \textit{delivery probability,} which
measures the capability of a MAC protocol to deliver the measure of
any sensor in the network to the intended destination (i.e., fusion
center, FC) and the \textit{time efficiency}\textbf{, }which measures
the data collection rate at the FC. The analysis is focused on standard
MAC protocols, such as TDMA, Framed-ALOHA (FA) and Dynamic-FA (DFA).
Novel design issues are also discussed, such as backlog estimation
and frame length selection. Extensive numerical results and discussions
validate the proposed analytical framework and provide insight into
the design of EH-WSNs.

\appendices{}

\section{Average Number of Sensor Transmissions per Time-slot\label{App: AVGnumberOfSensors}}

The conditional averages $\beta_{D,k}$ and $\beta_{C,k}$ are calculated
similarly to \cite{art: Schoute} by accounting for the capture effect
and an arbitrary $\rho$. Let $Y$ be the number of simultaneous transmissions
in the same slot, and let $\mathcal{U}_{k}$ and $\mathcal{C}_{k}$
respectively be the event of successful and collided slot in frame
$k$, the average number of sensors per successful and collided slot
are respectively 
\begin{equation}
\beta_{D,k}=\sum_{j=1}^{\infty}j\Pr\left[Y=j|\mathcal{U}_{k}\right];\text{ }\beta_{C,k}=\sum_{j=2}^{\infty}j\Pr\left[Y=j|\mathcal{C}_{k}\right]
\end{equation}
To calculate $\beta_{D,k}$ consider $\mathcal{A}.1$ and $\mathcal{A}.2$
and allow the number of possible interfering users up to infinity
as in Sec. \ref{Sec: Delivery_probability_FA}. By exploiting the
Bayes rule, we have $\Pr\left[Y=j|\mathcal{U}_{k}\right]=\Pr\left[\mathcal{U}_{k}|Y=j\right]\frac{\Pr\left[Y=j\right]}{\Pr\left[\mathcal{U}_{k}\right]}$,
where $\Pr\left[\mathcal{U}_{k}|Y=j\right]=jp_{c,k}(j-1)$, $\Pr\left[Y=j\right]=e^{-\frac{1}{\rho}}/(\rho^{j}j!)$
and $\Pr\left[\mathcal{U}_{k}\right]=p_{t,k}^{DFA}$ (see \ref{eqn: Time_efficiency_DFA_conditional}).
We can similarly obtain $\beta_{C,k}$ given that $\Pr\left[\mathcal{C}_{k}\right]=1-\Pr\left[\mathcal{U}_{k}\right]-\beta\left(0,B,L\right)$,
where $\beta\left(0,B,L\right)\simeq e^{-\frac{1}{\rho}}$ is the
probability of an empty slot, and $\Pr\left[\mathcal{C}_{k}|Y=j\right]=1-\Pr\left[\mathcal{U}_{k}|Y=j\right]$
for $j\geq1$.

\end{document}